# Automated Vehicle Crash Sequences: Patterns and Potential Uses in Safety Testing

Preprint[1]

Yu Song, Madhav V. Chitturi, David A. Noyce

Traffic Operations and Safety Laboratory, Department of Civil and Environmental Engineering
University of Wisconsin-Madison
1415 Engineering Dr, Madison, WI 53706

Please address correspondence to Yu Song: yu.song@wisc.edu

**Abstract**

With safety being one of the primary motivations for developing automated vehicles (AVs), extensive field and simulation tests are being carried out to ensure AVs can operate safely on roadways. Since 2014, the California Department of Motor Vehicles (DMV) has been collecting AV collision and disengagement reports, which are valuable data sources for studying AV crash patterns. A crash sequence of events describes the AV's interactions with other road users before a collision in a temporal manner. In this study, sequence of events data extracted from California AV collision reports were used to investigate patterns and how they may be used to develop AV test scenarios. Employing sequence analysis methods and clustering, this study evaluated 168 AV crashes (with AV in automatic driving mode before disengagement or collision) reported to the California DMV from 2015 to 2019. Analysis of subsequences showed that the most representative pattern in AV crashes was "collision following AV stop". Analysis of event transition showed that disengagement, as an event in 24% of all studied AV crash sequences, had a transition probability of 68% to an immediate collision. Cluster analysis characterized AV crash sequences into seven groups with distinctive crash dynamic features. Cross-tabulation analysis showed that sequence groups were significantly associated with variables measuring crash outcomes and describing environmental conditions. Crash sequences are useful for developing AV test scenarios. Based on the findings, a scenario-based AV safety testing framework was proposed with sequence of events embedded as a core component.

**Keywords:** Automated vehicle, Crash, Sequence, Disengagement, Test scenario

# 1    Introduction

Improving traffic safety is one of the primary motivations for developing automated vehicles (AVs). Apart from safer roads, AVs are predicted to bring other potential benefits such as improved mobility, better accessibility, lower energy consumption, and more efficient supply chains (NSTC and USDOT, 2020). Safety is prioritized as the top U.S. Government Automated Vehicle Technology Principle (NSTC and USDOT, 2020). Private and public organizations are taking efforts to ensure AV safety by extensively testing the vehicles on both closed courses and public roads (Koopman and Wagner, 2018; NHTSA, 2020a). In 2019, more than 1,400 automated vehicles were tested by more than 80 organizations across 36 U.S. states and Washington, D.C. (Etherington, 2019).

Since 2014, the California Department of Motor Vehicles (DMV) has required all permit-holding organizations that test AVs on California public roads to submit AV collision reports (California DMV, 2018). Prior to January 2020, two hundred and thirty-three (233) AV crashes were reported. In 168 of the reported cases, the AV was in automatic driving mode before disengagement or collision.

AV crash reports are a valuable data source for research to understand AV crash patterns. Prior explorations of California AV crashes provided some insights in:

- AV crash distribution by features such as manner of collision, AV-testing organization, year, and time of day
- Contributing factors of AV crashes and disengagements
- AV safety performance, measured by crash frequency per unit distance, compared with conventional (human-driven) vehicles

Crash information reported from AV field testing is useful in AV test scenario design. For example, Waymo is using data from field testing in developing challenging scenarios for closed-course and simulation-based AV testing (Schwall et al., 2020).

AV collision reports provide much more information than what has been used in previous studies. Sequence of events, which can be extracted from the crash report narratives, consists of information of chronologically ordered events happened in the crash. Many analytical methods commonly used in studying genome sequences can be used to characterize crash sequences. Compared with summarizing crashes with manners of collision or contributing factors, a crash characterization based on sequence of events better captures the crash progression characteristics. Differences in events or actions, and the order of events or actions can lead to different crash outcomes (Wu et al., 2016; Wu and Thor, 2015).

The primary objective of this study is to investigate the patterns in sequence of events in AV crashes. The secondary objective of this study is to discuss potential uses of crash sequences in scenario design for AV safety testing. Sequences of events were extracted from 168 AV collision reports' text narratives and analyzed using sequence analysis methods. Crash sequences, in combination with variables describing crash outcomes and variables describing the environment, can be used to design abstract semantic scenarios (Nitsche et al., 2017; Sander and Lubbe, 2018; Sui et al., 2019). A scenario-based AV testing framework, with crash sequence embedded as a core component, is proposed at the end of this study.

The contributions of this study are two-fold. This study adds to existing literature on California AV crashes and provides new insights by investigating AV crashes using sequence of events analysis. Beyond empirical findings, this study points out the practical application of crash sequences with a discussion on their potential uses in AV test scenario design.

## 2 Literature Review

Previous explorations of California AV crashes and disengagements started from 2015, when only limited data was available for aggregated statistical analyses. With increased AV road tests in recent years, more crashes and disengagements were reported to the California DMV. More recent studies were able to provide insights into the relationship among crashes, disengagements, and contributing factors by finding patterns in AV crashes and disengagements. Various analytical methods were used in previous studies. Some examples are statistical summary and tests, regression, classification trees, hierarchical Bayesian modeling, text mining, and clustering.

### 2.1 Patterns in AV Crashes

AVs crashes occurred mostly in the counties of Santa Clara and San Francisco, since the major AV testing organizations Waymo and Cruise carry out their AV testing in those two counties, respectively (Boggs et al., 2020b; Das et al., 2020; Favarò et al., 2017). AVs are tested on various types of roads including freeways/expressways, arterials, collectors, and local roads. AV crashes occurred on all roadway functional classes, with most crashes (60%) on arterial roads (Boggs et al., 2020b). Intersections are hotspots for AV crashes (Alambeigi et al., 2020; Banerjee et al., 2018; Boggs et al., 2020b; Favarò et al., 2017; Wang and Li, 2019a). Rear-end crashes were found to be the most common (60%) type of AV crashes (Alambeigi et al., 2020; Boggs et al., 2020b; Das et al., 2020; Dixit et al., 2016; Favarò et al., 2017; Leilabadi and Schmidt, 2019; Wang and Li, 2019a). Most (60%-80%) AV crashes occurred at a low relative speed between the AV and a second-party vehicle, usually below 10 mph (Banerjee et al., 2018; Favarò et al., 2017). The number of AV crashes is positively correlated with testing mileage, both periodically and cumulatively (Dixit et al., 2016; Favarò et al., 2017).

Two studies used clustering techniques to group AV crashes (Alambeigi et al., 2020; Das et al., 2020). Alambeigi et al. grouped 167 AV crashes based on themes in the description section of AV collision reports (Alambeigi et al., 2020). Alambeigi et al. identified five themes:

- Driver-initiated transition crashes
- Sideswipe crashes during left-side overtaking
- Rear-end crashes with vehicle stopped at an intersection
- Rear-end crashes with vehicle in a turn lane
- Crashes with oncoming traffic

Das et al. grouped 151 AV crashes into six clusters, based on crash attributes provided directly by the collision reports (Das et al., 2020). These clusters are:

- Two-vehicle non-injury crashes with unknown values in multiple attributes
- Single-vehicle non-injury crashes with unknown values in multiple attributes
- Injury crashes under poor lighting conditions during turning maneuvers or straight movement
- Single-case outlier cluster
- Two and multi-vehicle crashes with unknown values in multiple attributes
- Crashes with AV stopped and adverse weather conditions

### 2.2 Patterns in AV Disengagements

AV disengagements and test mileage data were aggregated and analyzed in previous studies. The cumulative disengagement number was positively correlated with cumulative test mileage (Banerjee et al., 2018; Favarò et al., 2018). Based on the mode of initiation, disengagements were classified into three types: automated, manual, and planned. Disengagement types were based on whether the disengagement was initiated by AVs, test operators, or as a part of a planned fault injection campaign (Banerjee et al., 2018; Dixit et al., 2016; Favarò et al., 2017). Before 2018, not all AV testing organizations reported disengagements with a clear differentiation between initiation modes. In a 2014-2016 sample of Waymo reported disengagements, about half were manually initiated, and the other half were automatically initiated (Banerjee et al., 2018). In the most recent study of California AV disengagements by Boggs et al., 25% of the sampled disengagements were human initiated (Boggs et al., 2020a). The monthly automatic disengagement number was highly correlated with the monthly manual disengagement number, indicating test operators' trust in the AVs' capability to navigate through risks (Dixit et al., 2016).

In terms of factors causing disengagement, the California DMV does not provide predefined categories, so different categorizations were used in previous studies to analyze the data (Banerjee et al., 2018; Boggs et al., 2020a; Dixit et al., 2016; Favarò et al., 2018; Lv et al., 2018; Wang and Li, 2019a). Overall, system issues, including AVs' perception, planning, and decision-making, caused over half of the disengagements. In the Boggs et al. study, system issues were reported to have caused 89% of the disengagements, with a breakdown into control discrepancy (7%), hardware and software discrepancy (26%), perception discrepancy (21%), and planning discrepancy (35%) (Boggs et al., 2020a).

Disengagement reaction time was available in the 2015-2017 disengagement reports, and was another important information studied in past literature (Banerjee et al., 2018; Dinges and Durisek, 2019; Dixit et al., 2016). The California DMV defined disengagement reaction time as "the period of time elapsed from when the autonomous vehicle test driver was alerted of the technology failure, and the driver assumed manual control of the vehicle" (Banerjee et al., 2018). The average disengagement reaction time was estimated to be 0.83 s – 0.87 s (Banerjee et al., 2018; Dinges and Durisek, 2019; Dixit et al., 2016). The average disengagement reaction time is between the average automobile brake reaction time of 1.13 s and the average motorcycle brake reaction time of 0.60 s (Dinges and Durisek, 2019). Disengagement reaction time is expected to increase with test operators becoming more comfortable and gaining trust in the AV's handling of risky situations on roads (Banerjee et al., 2018; Dinges and Durisek, 2019).

**2.3    Relationship Among AV Crashes, Disengagements, and Contributing Factors**

Many more disengagements than AV crashes have occurred with some disengagements followed by crashes. Banerjee et al. found in the 2014-2016 data that 23% of AV crashes involved disengagements, but a very small fraction (0.8%) of AV disengagements led to crashes (Banerjee et al., 2018). Favarò et al., using the 2014-2017 data, found that 1 in every 178 (0.5%) disengagements led to a crash (Favarò et al., 2018).

Previous studies used various statistical modeling methods to evaluate the relationship between contributing factors and AV crash outcomes described by crash type, manner of collision, and severity (Boggs et al., 2020b; Leilabadi and Schmidt, 2019; Wang and Li, 2019b). Leilabadi and Schmidt found that adverse road surface conditions were significantly associated with a higher severity of AV damage. Also, 80% of the crashes when AV was in automated driving mode were identified as hit-and-run crashes, in which 40% were rear-end and 40% were sideswipe crashes. Wang and Li explored the mechanism of contributing factors affecting AV crash severity and manner of collision, and found that:

- Injury crashes (study included cases from outside California) happened when an AV was in automated driving mode and was responsible for the crashes
- All intersection crashes were rear-end, while roadway segment crashes with AVs in automated driving mode were angle or sideswipe

Boggs et al. focused on the factors affecting rear-end AV crashes and injury AV crashes, and found that:

- AVs in automated driving mode were more likely to get involved in rear-end crashes (without specifying if AVs were rear-ended or AVs rear-ended others) than AVs in manual driving mode or were disengaged from automatic driving mode
- In a mixed land use environment, AV crashes were more likely to be rear-end
- Higher speeds of second-party vehicles, no marked centerline, and non-clear weather were more likely to be associated with injury AV crashes

Two studies evaluated the relationship between various contributing factors and features of AV disengagements (Boggs et al., 2020a; Wang and Li, 2019a). Wang and Li explored factors leading to disengagements in different stages of an AV's operation (perception, planning, and control phases) and disengagements with different take-over time (divided into two groups with a threshold of 0.5 s). Major findings of Wang and Li were:

- For an AV, a larger number (> 5) of radar sensors and a more appropriate number (3-4) of LiDAR sensors would lead to fewer disengagements
- Disengagements on local roads and freeways were associated with a shorter take-over time

Boggs et al. focused on factors affecting the mode of disengagement initiation (manual or automatic), and found that:

- Planning discrepancy, software/hardware issues, and environmental/other road user issues all significantly increased the probability of an automatic initiation to different extents
- More automatic initiations occurred as time progressed month by month

**2.4   Safety Performance of AVs Compared with Conventional Vehicles**

AV crash records were compared with conventional vehicle (human-driven vehicle) crash records, by evaluating crash rates (in crashes per mile driven), injury rates (injuries per mile driven), crash types, and crash severity (Banerjee et al., 2018; Blanco et al., 2016; Schoettle and Sivak, 2015; Teoh and Kidd, 2017). Conventional vehicle crash records were obtained from databases such as the National Highway Traffic Safety Administration (NHTSA) Fatality Analysis Reporting System (FARS), National Automotive Sampling System-General Estimates System (NASS-GES), and the Federal Highway Administration (FHWA) Strategic Highway Research Program Naturalistic Driving Study (SHRP2 NDS) database. With a small sample size of AV crashes (fewer than 20 cases), comparative studies could not reach an agreement nor a definite conclusion on whether AVs perform better than conventional vehicles in terms of safety (Blanco et al., 2016; Schoettle and Sivak, 2015; Teoh and Kidd, 2017). In a more recent study by Banerjee et al. California AV collision reports from 2014-2016 (42 crashes) and NHTSA 2015 motor vehicle crash data were used for an AV-human driver performance comparison. Banerjee et al. claimed that current AVs are 15 to 4,000 times worse than human drivers in terms of safety performance measured by crashes per cumulative mile driven (Banerjee et al., 2018).

**2.5   Sequence of Events in Traffic Crashes**

The Tri-Level Study of the Causes of Traffic Accidents found that 50% of the 2,000 crashes studied were caused by more than one factor (Treat et al., 1979). Sequence of events is important information for traffic crash investigation, and was recommended by the National Transportation Safety Board (NTSB) to be included in national crash databases (NTSB, 2011). Crash progression patterns can be discovered through crash sequence analysis and are helpful in identifying effective prevention strategies (Wu and Thor, 2015). Sequence analysis was developed in bioinformatics to analyze genome sequences, and is also applied in social sciences (Abbott, 1995; Cornwell, 2015). Genome sequence analysis methods is applicable to sequence of events to study traffic crash patterns (Cornwell, 2015; Wu et al., 2016). Wu et al. used sequence analysis on Fatality Analysis Reporting System (FARS) data to group similar crashes and model crash severity outcomes (Wu et al., 2018, 2016). A similar type of crash sequence data can be extracted from the California AV collision reports and analyzed using sequence analysis methods. Through sequence analysis, frequent pre-collision events can be identified, the stochastic relationships between events can be evaluated, and whole sequences can be classified into types that represent distinctive crash progression characteristics. Sequence pattern information, together with other crash attributes and environmental (both man-made and natural) condition variables, can be used in designing representative AV test scenarios.

# 3 Data

## 3.1 AV Crashes

AV collision reports from 2014 to 2019 were obtained from the California DMV. "Report of Traffic Accident Involving an Autonomous Vehicle (OL 316)" was the required form for AV testing organizations to submit. The form was updated in 2017, adding information such as weather, lighting, and road surface conditions. All AV collision reports were archived by the California DMV and are publicly available online. A total of 233 reports were gathered and manually reviewed, with key information transferred into a spreadsheet. The 168 reports of crashes where the AV was in automatic driving mode before disengagement or collision were used for analysis in this study. Table 1 presents a summary of several data elements of those reports.

AV crashes in California increased each year during 2015-2019. Data show that 43% of AV crashes occurred in 2019, due to the fact that 43% (2.58 million) of AV testing miles were driven in 2019 during the five-year period. Over 76% of AV crashes happened during the months of May-November. Also, 71% of AV crashes happened during daytime (7:00 am – 6:00 pm). Ninety-five percent of the AV crashes took place in San Francisco, Mountain View, and Palo Alto, where most AV testing was carried out in California.

Most (98%) of the 168 AV crashes occurred between AVs and other road users, and 2% were single-vehicle crashes. AV crashes involving bicyclists, e-scooters, pedestrians, and skateboarders were 8% of all crashes. AV crashes involving motorcycles were 5% of all crashes. AV crashes mostly (73%) took place at intersections (including ramp terminals), followed by roadway segment (26%) or in a parking lot (2%). Intersections where AV crashes occurred were primarily signal controlled, as stated in 51% of AV collision reports.

Rear-end (62%) and sideswipe (21%) were the two most common manners of collision in AV crashes. Injuries, without differentiating between minor or serious, were reported in 12% of the crashes. Disengagements were reported in 24% of the crashes. In 35% of the cases, AVs were yielding to another road user before a collision took place. In more than half of the crash cases, AVs were stopped (41%) or slowing down (10%). AVs were proceeding straight in 32% and turning in 12% of the crashes. Second-party road users were proceeding straight in 55% and turning in 15% of the crashes.

Of the 168 AV crashes, 37% and 57% were reported by Waymo and Cruise, respectively. During 2015-2019, Waymo and Cruise ran the most (4.1 million) and second most (1.4 million) public-road AV testing miles (in automatic driving mode), respectively. Waymo and Cruise's AV mileages accounted for 67% and 23% of the total mileage (6.2 million) reported by all AV-testing organizations that reported AV crashes in 2015-2019. Table 2 lists AV test mileages and crash rates (per million miles) by organization, sorted by mileage in ascending order. Waymo and Cruise yielded crash rates of 15 and 67 crashes per million miles tested. Comparatively, the 2018 passenger car crash rate in United States was approximately 4.4 crashes per million vehicle miles traveled (NHTSA, 2020b). Note that the AV crash rates are not reflective of automatic driving systems' true safety performance, as there were still human operator interventions involved in the studied AV crashes. Without complete information from testing automatic driving systems without human intervention, we cannot make a true comparison between automatic driving systems and human drivers. However, the information from currently available AV crash reports is to some extent useful in measuring the evolution of AV technology.

## Table 1 Summary of data from California AV collision reports

| Field | Count | Percentage | Field | Count | Percentage |
|---|---|---|---|---|---|
| **Year** | | | **Month** | | |
| 2015 | 9 | 5% | 1 | 5 | 3% |
| 2016 | 12 | 7% | 2 | 10 | 6% |
| 2017 | 24 | 14% | 3 | 9 | 5% |
| 2018 | 50 | 30% | 4 | 8 | 5% |
| 2019 | 73 | 43% | 5 | 16 | 10% |
| **City** | | | 6 | 19 | 11% |
| Fremont | 1 | 1% | 7 | 19 | 11% |
| Los Altos | 5 | 3% | 8 | 19 | 11% |
| Milpitas | 1 | 1% | 9 | 14 | 8% |
| Mountain View | 42 | 25% | 10 | 25 | 15% |
| Palo Alto | 17 | 10% | 11 | 17 | 10% |
| San Francisco | 100 | 60% | 12 | 7 | 4% |
| Sunnyvale | 2 | 1% | **Time of Day** | | |
| **AV Testing Organization** | | | Day | 119 | 71% |
| Apple | 1 | 1% | n/a | 6 | 4% |
| Cruise | 95 | 57% | Night | 43 | 26% |
| Google Auto (Waymo) | 22 | 13% | **Second Party Type** | | |
| Jingchi (WeRide) | 1 | 1% | Bike | 6 | 4% |
| Lyft | 1 | 1% | Bus | 3 | 2% |
| Pony.AI | 2 | 1% | Car | 139 | 83% |
| UATC (Uber) | 1 | 1% | E-scooter | 3 | 2% |
| Waymo | 41 | 24% | Motorcycle | 8 | 5% |
| Zoox | 4 | 2% | n/a | 4 | 2% |
| **Facility Type** | | | Pedestrian | 2 | 1% |
| Road Segment | 43 | 26% | Skateboarder | 1 | 1% |
| Intersection | 122 | 73% | Truck | 1 | 1% |
| Parking lot | 3 | 2% | Van | 1 | 1% |
| **Traffic Control** | | | **AV Mode** | | |
| AWSC | 11 | 7% | Automatic | 127 | 76% |
| Crosswalk Sign | 1 | 1% | Automatic-Manual | 41 | 24% |
| n/a | 52 | 31% | **AV Yielding** | | |
| Signal | 86 | 51% | No | 109 | 65% |
| Stop Sign | 10 | 6% | Yes | 59 | 35% |
| TWSC | 1 | 1% | **Severity** | | |
| Xwalk Flashing Light | 1 | 1% | Injury | 20 | 12% |
| Yield Sign | 6 | 4% | Non-Injury | 148 | 88% |
| **Manner of Collision** | | | **Second Party Movement** | | |
| Broadside | 12 | 7% | Backing | 3 | 2% |
| Hit Object | 3 | 2% | Changing Lanes | 16 | 10% |
| Other | 3 | 2% | Entering Traffic | 2 | 1% |
| Rear End | 112 | 67% | Making Left Turn | 10 | 6% |
| Sideswipe | 37 | 22% | Making Right Turn | 15 | 9% |
| Vehicle/Pedestrian | 1 | 1% | Merging | 5 | 3% |
| **AV Movement** | | | n/a | 6 | 4% |
| Changing Lanes | 8 | 5% | Other | 2 | 1% |
| Making Left Turn | 11 | 7% | Other Unsafe Turning | 2 | 1% |
| Making Right Turn | 8 | 5% | Parked | 1 | 1% |
| Merging | 1 | 1% | Passing Other Vehicle | 8 | 5% |
| Passing Other Vehicle | 1 | 1% | Proceeding Straight | 92 | 55% |
| Proceeding Straight | 53 | 32% | Slowing/Stopping | 6 | 4% |
| Slowing/Stopping | 17 | 10% | | | |
| Stopped | 69 | 41% | | | |

Table 2  2015-2019 AV test mileages and crash rates by organization

| Organization | Test Mileage | Mileage Share | Crashes | Crashes per Million Miles |
|---|---|---|---|---|
| Waymo | 4,122,878 | 68.6% | 63 | 15 |
| Cruise | 1,420,360 | 23.6% | 95 | 67 |
| Pony.AI | 192,642 | 3.2% | 2 | 10 |
| Zoox | 100,023 | 1.7% | 4 | 40 |
| Apple | 88,283 | 1.5% | 1 | 11 |
| Lyft | 42,931 | 0.7% | 1 | 23 |
| UATC (Uber) | 26,899 | 0.4% | 1 | 37 |
| Jingchi (WeRide) | 19,067 | 0.3% | 1 | 52 |
| *All* | *6,013,083* | *100.0%* | *168* | *28* |

### 3.2   AV Disengagements in Crashes

AV disengagement reports from 2015 to 2019 were obtained from California DMV. "Annual Report of Autonomous Vehicle Disengagement (OL 311R)" was the required form for AV-testing organizations to submit. The form was updated in 2017, before which there was no uniform format for disengagement reporting. Disengagement reports provide information such as a summary of AV test mileages, number of disengagements, disengagement dates, locations (highway or street), and a description of disengagement causes.

Information from the disengagement reports were matched to the AV crash records which involved disengagements. A breakdown of different causes for AV disengagements followed by a crash is shown in Table 3. Categories of causes were created based on our comprehension of different descriptions provided by different AV testing organizations' reports, as uniform terms for describing disengagement causes were not provided and are not required by the California DMV. Not all disengagements as reported in AV collision reports were reported in AV disengagement reports. For such cases, causes were summarized based on their collision report descriptions.

Table 3  Causes of disengagements involved in AV crashes

| Disengagement Cause | Count | Percentage |
|---|---|---|
| Operator precaution | 19 | 46% |
| Reckless road user | 16 | 39% |
| Unwanted movement | 3 | 7% |
| Planned | 2 | 5% |
| Operator error | 1 | 2% |
| *Total* | *41* | *100%* |

Of the 41 disengagements in AV crashes, 19 (46%) were initiated by an operator out of precaution, and 16 (39%) were a reaction to a nearby reckless road user. The rest of the disengagements were initiated due to unwanted AV movements (3, 7%), operator error (1, 2%), or were planned tasks for tests (2, 5%). None of the 41 disengagements were caused by vehicle system (perception, hardware or software) problems. It could not be determined whether or not all reckless-road-user-caused disengagements were initiated by an operator or an AV itself, as the information was not clearly stated in all AV disengagement and collision reports.

### 3.3   AV Crash Sequences

In this study, an AV crash sequence consists of events. An event is an action or a collision. A collision can happen between an AV and an object, or between an AV and another road user. In some crash cases, multiple road users were involved. For crash sequences used in this study, AV was denoted as "v1"; a second-party road user (vehicle, bicyclist, pedestrian, or others) that collided with the AVs, was denoted as "v2"; and a third-party road user that interacted (but may or may not have collided) with the AVs or second-party road users, was denoted as "v3".

Events were extracted from text narratives in the AV collision reports, which were reviewed and summarized manually. AV crash sequence lengths ranged from 2 to 5 events, with an average of 2.8. Events were first recorded using short phrases such as "v1 stop" and "v2 pass v1 on right". Ordering of events in sequences was based on temporal information provided by the text. When recording events, consistency was maintained in the use of short phrases. Each short phrase went through a second round of review and was encoded with a label, which was a combination of English alphabet letters and/or Arabic numerals. To further enhance consistency, phrases describing similar events (based on our understanding of traffic crashes and judgement) were encoded with the same label. For example, "v2 run stop sign" and "v2 run red light" were encoded with the same label, "V2", since both phrases describe a second-party road user's violation of traffic control at an intersection. Following this procedure of "text narratives → short phrases → labels", we converged to a set of 35 labels for the encoding of 497 events, which made up the 168 AV crash sequences. Of the 35 different labels, 14 denoted actions initiated by AVs; 14 denoted actions initiated by second-party road users/objects; and 7 denoted actions initiated by third-party road users. Detailed encodings are listed in Table 4.

Table 4  Event encoding

| Label | Short Phrase | Count | Label | Short Phrase | Count |
|---|---|---|---|---|---|
| A1 | v1 accelerate/proceed | 39 | PL1 | v1 pass v3 on left | 1 |
| A2 | v2 accelerate/proceed | 2 | PL2 | v2 pass v1 on left | 13 |
| B1 | v1 back up | 1 | PL3 | v3 pass v1 on left | 1 |
| B2 | v2 back up | 1 | PR2 | v2 pass v1 on right | 7 |
| D1 | v1 decelerate | 32 | PR3 | v3 pass v1 on right | 1 |
| D2 | v2 decelerate | 1 | R1 | v1 make right turn | 9 |
| D3 | v3 decelerate | 1 | R2 | v2 make right turn | 2 |
| DG | v1 disengage | 41 | S1 | v1 stop | 76 |
| DT | v1 detect v2 | 2 | S2 | v2 stop | 2 |
| L1 | v1 make left turn | 11 | SA2 | v2 stop and proceed | 2 |
| L2 | v2 make left turn | 9 | V2 | v2 run stop sign/red light | 5 |
| L3 | v3 make left turn | 1 | X12 | v1 contact v2 | 7 |
| ML1 | v1 merge left | 6 | X10 | v1 hit object | 3 |
| ML2 | v2 merge left | 16 | X21 | v2 contact v1 | 155 |
| ML3 | v3 merge left | 6 | X32 | v3 contact v2 | 1 |
| MR1 | v1 merge right | 7 | XO1 | object/person contact v1 | 3 |
| MR2 | v2 merge right | 12 | Y | v1 yield | 15 |
| MR3 | v3 merge right | 6 | | | |

Table 5 gives an example of two crash event sequences. A sequence consists of one or more elements, each of which represents a pre-collision or collision event. A subsequence is a set of chronologically ordered elements (following the order in sequence but not necessarily adjacent) that appears in a larger sequence (Cornwell, 2015). A subsequence that consists of consecutive elements is called a substring, or an n-gram, with n referring to the number of elements in the subsequence. For example, Sequence 1 and Sequence 2 in Table 5 both have a subsequence "S1-X21", with two elements, "S1" and "X21". Sequence 1 and Sequence 2 both have a substring, or 2-gram, "PR2-X21", with two elements, "PR2" and "X21". Elements, subsequences, and whole sequences were all analyzed in this study to understand patterns in AV crashes.

Table 5  Example of crash event sequences

| Sequence | Element 1 | Element 2 | Element 3 | Element 4 |
|---|---|---|---|---|
| Sequence 1 | S1 | PR2 | X21 | |
| Sequence 2 | S1 | A1 | PR2 | X21 |

## 4  Methodology

As mentioned, the primary objective of this study was to identify AV crash sequence patterns. Sequences were analyzed at three levels: the element level, subsequence level, and whole sequence level. Element-level analysis investigated the basic components of the sequence and the components' weight in the entire element space. Subsequence-level analysis investigated the stochastic relationships between elements. Whole-sequence-level analysis investigated the similarities and dissimilarities between sequences, which were used to identify groups or classes of sequences. In the context of a traffic crash sequence study, we were interested in identifying frequently occurring events, quantifying the interconnections between events, as well as classifying crash progressions. In this study, AV crash sequence analysis focused on identifying patterns from these three aspects: 1) most frequent events and event transitions; 2) disengagements' role in AV crash sequences; and 3) characterization of AV crash sequences.

Descriptive analysis was used to summarize frequencies of events and subsequences in AV crash sequences. Stochastic patterns in event transitions were presented by a transition matrix. To characterize crash sequences, a cluster analysis was carried out.

Following the sequence analysis, a discussion is presented at the end of this paper, about potential uses of AV crash sequences in scenario-based AV safety testing. In the discussion, a cross-tabulation analysis was used between crash sequence groups, other AV crash attributes, and environmental condition variables. In this section, concepts and methods used in AV crash sequence analysis are introduced in detail.

### 4.1 Transition Matrix

A transition matrix shows the probability of transition between every pair of adjacent positions in all sequences (Cornwell, 2015). The size of a transition matrix is k × k, where k is the number of elements in the element universe. The rows of a transition matrix indicate elements where transitions are from, and the columns indicate elements where transitions are to. A transition matrix, denoted as P, has the form shown in Figure 1. Cell $P_{AB}$ contains the probability, p(AB), that element A is followed by element B in all cases that element A appears in the element universe, which is calculated as:

$$p(AB) = p(B_p|A_{p-1}) = \frac{n(AB)}{n(A)} \qquad [1]$$

where n(AB) = number of times that 2-gram AB appears; and

n(A) = number of times that element A appears.

|  | Element at position *p* | | |
|---|---|---|---|
|  | A | B | C |
| Element at position *p-1*  A | p(AA) | p(AB) | p(AC) |
| B | p(BA) | p(BB) | p(BC) |
| C | p(CA) | p(CB) | p(CC) |

**Figure 1  Form of transition matrix P (Cornwell, 2015)**

The sum of probabilities in each row is 1. Note that p(AB) does not capture both the probability $p(B_p|A_{p-1})$ (the conditional probability that B appears given that A has just appeared) and $p(A_{p-1}|B_p)$ (the conditional probability that A appeared just before given that B appears) (Cornwell, 2015). Transition matrix P does not show $p(A_{p-1}|B_p)$, which should be calculated as:

$$p(A_{p-1}|B_p) = \frac{n(AB)}{n(B)} \qquad [2]$$

where n(B) = number of times that element B appears.

## 4.2 Measuring Sequence Dissimilarity and Optimal Matching

To compare and group AV crashes based on sequences, we need to measure the dissimilarity (or distance) between sequences. A common approach for sequence comparison is optimal matching (OM), which is widely used in genome sequence and social sequence analysis (Cornwell, 2015; Kruskal, 1983).

When comparing two sequences, the distance between them is defined by the "cost" to transform a sequence to the other. This transformation is called "alignment", and "cost" is measured by the number of different operations needed to complete the alignment. There are multiple ways to align two sequences. For example, Table 6 shows two of the multiple ways that can be taken to align the two sequences in Table 5. The operations applied include insertion, deletion, and substitution. The cost of insertion or deletion is denoted by "d" and is called "indel" cost. The cost of substitution is denoted by "s". There are other more intricate operations for sequence alignment, but indels and substitutions are commonly used and have been previously used for traffic crash sequence analysis (Wu et al., 2016). The sum of cost for sequence alignment is a measure of dissimilarity (or distance) between two sequences. In our Table 6 example, the first alignment method uses both indels and substitutions and costs 2s+d, while the second alignment method only uses indels and costs d. Depending on the selection of operations and setting of operation costs, there are different sequence distance metrics (Cornwell, 2015). Three basic and commonly used ones are:

- Levenshtein distance that uses both indels and substitutions, with the same cost of 1
- Levenshtein II distance that uses only indels with a cost of 1
- Hamming distance that only considers substitutions

Operation costs can also be set based on the needs of analysis and the properties of sequences. In previous analysis of traffic crash sequences, the Levenshtein distance was used as the measure of dissimilarity (Wu et al., 2016). In this analysis of AV crash sequences also, the Levenshtein distance was used.

**Table 6  Example of ways to align two sequences**

| Sequence 1 | S1 | PR2 | X21 | | |
| Sequence 2 | S1 | A1  | PR2 | X21 | |
| **Alignment 1** | | | | | |
| Sequence 1 | S1 | PR2 | X21 | ø | |
| Sequence 2 | S1 | A1  | PR2 | X21 | |
| *Cost* | 0 | s | s | d | = 2s+d |
| **Alignment 2** | | | | | |
| Sequence 1 | S1 |     | PR2 | X21 | |
| Sequence 2 | S1 | A̶1̶ | PR2 | X21 | |
| *Cost* | 0 | d | 0 | 0 | = d |

Note: Insertion is marked with ø;
Deletion is marked with s̶t̶r̶i̶k̶e̶t̶h̶r̶o̶u̶g̶h̶; and
Substitution is marked with underline.

As there can be multiple ways of aligning a pair of sequences which generate different distance values, the alignment that generates the smallest distance value should be found, and the smallest distance should be used as the measure of dissimilarity between that pair of sequences (Cornwell, 2015; Wu et al., 2016). An OM procedure finds the dissimilarity between every pair of sequences in a sequence space. The Needleman-Wunsch algorithm is a classic OM algorithm which is widely used in bioinformatics to align sequences and find sequence dissimilarities (Needleman and Wunsch, 1970). For two sequences, A and B, an empty matrix L, of size length(A)+1 by length(B)+1 is created. Based on a set of indel and substitution costs (e.g., for Levenshtein distance, indel and substitution costs are both 1), the Needleman-Wunsch algorithm fills matrix L and returns the smallest alignment cost (distance) between sequences A and B. Pseudocode of the Needleman-Wunsch algorithm is as follows (Needleman and Wunsch, 1970; Wu et al., 2016).

```
Algorithm Needleman-Wunsch(A, B)
```

```
# initialize
L <- matrix of size length(A)+1 * length(B)+1
d <- indel cost
s <- substitution cost
# fill the cells of L
for i = 0 to length(A)
        L(i,0) <- d*i
for j = 0 to length(B)
        L(0, j) <- d*j
for i = 1 to length(A)
        for j = 1 to length(B) {
                insert <- L(i, j−1) + d
                delete <- L(i−1, j) + d
                substitute <- L(i−1, j−1) + s
                L(i, j) <- max(insert, delete, substitute)
                }
# smallest alignment cost (distance)
return L(length(A), length(B))
```

The dissimilarity between every pair of sequences in the studied sample was calculated using the Needleman-Wunsch algorithm. A dissimilarity matrix was formed and used in cluster analysis as the basis for clustering similar sequences.

### 4.3 Cluster Analysis

There are various cluster analysis methods and different ones can produce different clustering results. The k-medoids method was selected for this sequence clustering because the k-medoids method works well with categorical data (such as sequences) and is robust against outliers (Nitsche et al., 2017). Most commonly, the k-medoids method is implemented by the partitioning around medoids (PAM) algorithm developed by Kaufmann and Rousseeuw (Kaufmann and Rousseeuw, 1987). The PAM algorithm asks for the number of demanded clusters, k (k ≤ sample size), and greedily finds k points from the sample set (denoted as X) as medoids (denoted as M) to form clusters. In this analysis, X is in the form of a dissimilarity matrix. The objective of the algorithm is to minimize a cost measured by the sum of distances between each x (∈ X) to its assigned cluster medoid m (∈ M). Pseudocode of PAM algorithm is as follows (Kaufmann and Rousseeuw, 1987; Nitsche et al., 2017; Phillips, 2019).

```
Algorithm PAM(X, k)
```

```
# build
choose k points M ⊂ X
for all x ∈ X
        assign x to X_i if x is closest to m_i
calculate cost
# swap
repeat
for all m ∈ M
        for all x ∈ X and ∉ M {
                swap x with m; calculate cost
                if (cost decreases) keep x and m
                else do not swap
                }
until (cost does not change)
```

The PAM algorithm was applied to the 168 AV crash sequences, with the k value ranging from 2 to 10. A measure for evaluating quality of clustering used for this analysis is called the "silhouette", which describes how well a data point lies within its own cluster compared to other clusters (Rousseeuw, 1987). Silhouette width (value) is calculated as (Rousseeuw, 1987):

$$s_i = \frac{b_i - a_i}{\max\{a_i, b_i\}}$$

where $s_i$ = silhouette width;

$a_i$ = average dissimilarity of object i to all other objects of A (the cluster that i is assigned to); and

$b_i = \min\limits_{C \neq A} d_{i,C}$, with $d_{i,C}$ = average dissimilarity of i to all objects of C (any cluster that i is not assigned to).

Silhouette width is between -1 and 1, with a higher value meaning a better clustering. When there is only one object in a cluster, the object's silhouette width is 0.

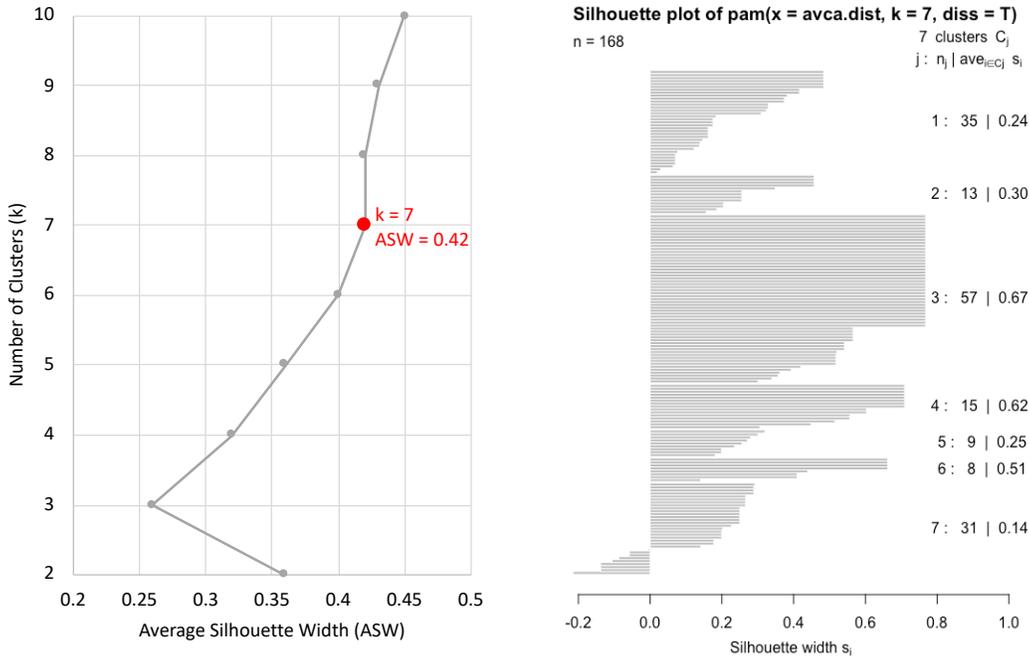

(a) Average silhouette widths, k = 2 to 10        (b) Silhouette widths, k = 7

**Figure 2  Silhouette widths**

**Table 7  Cluster size and cluster average silhouette width**

| Number of Clusters: "k" | Cluster Size | | Cluster Avg. Silhouette Width | | Avg. Silhouette Width |
| --- | --- | --- | --- | --- | --- |
| | Min | Max | Min | Max | |
| 2 | 51 | 117 | 0.14 | 0.45 | 0.36 |
| 3 | 48 | 63 | -0.05 | 0.74 | 0.26 |
| 4 | 15 | 57 | -0.03 | 0.74 | 0.32 |
| 5 | 8 | 57 | 0.02 | 0.73 | 0.36 |
| 6 | 8 | 57 | 0.04 | 0.62 | 0.4 |
| 7 | 8 | 57 | 0.14 | 0.67 | 0.42 |
| 8 | 7 | 55 | 0.11 | 0.7 | 0.42 |
| 9 | 7 | 55 | 0.16 | 0.69 | 0.43 |
| 10 | 5 | 53 | 0.11 | 0.73 | 0.45 |

The average silhouette widths from sequence clustering with different k values are plotted in Figure 2(a). The most appropriate k values were selected based on an evaluation of both overall and cluster-wise average silhouette widths. The overall average silhouette width should be as close to 1 as possible, with each cluster's silhouette width being larger than 0.1. Also, k should be preferably small, for easy cluster interpretation and to avoid "overfitting". Changes in cluster size and cluster average silhouette width are shown in Table 7. When k = 7, we could obtain a relatively small number of clusters, with a large enough average silhouette width and better cluster average silhouette widths than obtained using other k values. Therefore, k = 7 was used. Detailed silhouette plot for clustering with k = 7 is shown in Figure 2(b).

## 5 Results

### 5.1 Most Frequent Subsequences

To grasp overall patterns in crash sequences, we investigated the 15 most representative subsequences, which are listed in Table 8. The results showed that 92% of AV crash sequences ended with AV hit by a second-party road user. In 40% of the crash sequences, the AV stopped and was hit by a second-party road user. Disengagement was an event in 24% of the AV crash sequences. AV hit by a second-party road user following disengagement appeared in 19% of the crash sequences. Colliding right after the AV started moving (21%) or the AV started decelerating (19%) are two other common subsequences, which indicates crash cases where AVs were possibly violating expectancies of other road users. Other top-15 subsequences include AVs' yielding and second-party road users' merging actions.

**Table 8  Top 15 most frequent subsequences**

| Rank | Subsequence | Description | Count | % |
|---|---|---|---|---|
| 1 | (X21) | (AV hit by 2$^{nd}$ party) | 155 | 92% |
| 2 | (S1) | (AV stops) | 71 | 42% |
| 3 | (S1)-(X21) | (AV stops) then (AV hit by 2$^{nd}$ party) | 68 | 40% |
| 4 | (DG) | (AV disengaged) | 41 | 24% |
| 5 | (A1) | (AV accelerates/proceeds) | 38 | 23% |
| 6 | (A1)-(X21) | (AV accelerates/proceeds) then (AV hit by 2$^{nd}$ party) | 36 | 21% |
| 7 | (D1) | (AV decelerates) | 32 | 19% |
| 8 | (D1)-(X21) | (AV decelerates) then (AV hit by 2$^{nd}$ party) | 32 | 19% |
| 9 | (DG)-(X21) | (AV disengaged) then (AV hit by 2$^{nd}$ party) | 32 | 19% |
| 10 | (D1)-(DG) | (AV decelerates) then (AV disengaged) | 16 | 10% |
| 11 | (D1)-(DG)-(X21) | (AV decelerates) then (AV disengaged) then (AV hit by 2$^{nd}$ party) | 16 | 10% |
| 12 | (ML2) | (2$^{nd}$ party merges left) | 16 | 10% |
| 13 | (ML2)-(X21) | (2$^{nd}$ party merges left) then (AV hit by 2$^{nd}$ party) | 15 | 9% |
| 14 | (Y) | (AV yields) | 15 | 9% |
| 15 | (Y)-(X21) | (AV yields) then (AV hit by 2$^{nd}$ party) | 15 | 9% |

### 5.2 Transitions to and from Disengagement

A 35 by 35 transition matrix was obtained. Since transitions to and from disengagements (DG) were of the most interest, relevant results are illustrated in Figure 3. Note that transition rates in the left column do not add up to 100%, but the ones in the right column add up to 100%, because of the reason explained previously in the Transition Matrix part of the Methodology section. Figure 3 helps identify the preceding and succeeding events of AV disengagements. Disengagements were initiated generally in two types of situations:

- AVs or human operators detected reckless actions of second or third-party road users
- Human operator felt uncomfortable with some driving maneuvers made by AVs

Examples of possible reactions to reckless actions are that 100% of "second-party road user decelerates" (D2) events were followed by AV disengagements (DG), and 50% of "third-party road user merging right" (MR3) events were followed by AV disengagements (DG). Possibly out of operators' discomfort with AV's actions, 44% of "AV deceleration" (D1) events and 43% of "AV merging right" (MR1) events were followed by disengagements.

While none of the studied disengagements were able to help avoid collisions (as these disengagements were all in crash sequences), 68% of them were followed by an immediate collision rather than being followed by certain other actions before collisions. Immediately after 51% of disengagement events, second-party road users hit the AVs. Following 10% and 7% of disengagement events, the AVs hit second-party vehicles or objects, respectively. In the other 32% of cases, there was still time for AVs or second-party road users to take some actions before the collision.

|  | AV disengaged | | |
|---|---|---|---|
|  | [-> DG] | [DG ->] |  |
| AV decelerates [D1 ->] | 44% | 51% | [-> X21] 2nd party hits AV |
| 2nd party decelerates [D2 ->] | 100% | 10% | [-> X12] AV hits 2nd party |
| 2nd party makes left turn [ML2 ->] | 31% | 7% | [-> X1O] AV hits objects |
| 3rd party makes left turn [ML3 ->] | 33% | 7% | [-> MR1] AV merges right |
| AV merges right [MR1 ->] | 43% | 5% | [-> D1] AV decelerates |
| 2nd party merges right [MR2 ->] | 17% | 5% | [-> ML2] 2nd party merges left |
| 3rd party merges right [MR3 ->] | 50% | 5% | [-> V2] 2nd party runs stop sign/red light |
| 2nd party passes AV from left [PL2 ->] | 15% | 2% | [-> B1] AV backs up |
| AV stops [S1 ->] | 1% | 2% | [-> L2] 2nd party makes left turn |
| 2nd party runs stop sign/red light [V2 ->] | 20% | 2% | [-> ML1] AV merges left |
| AV yields [Y ->] | 27% | 2% | [-> MR2] 2nd party merges right |

**Figure 3 Transition rates from preceding events to disengagement and from disengagement to succeeding events**

## 5.3 Sequence Characterization

Cluster analysis resulted in crash sequences being clustered into 7 groups, as shown in Table 9. Patterns in crash sequences within each group and differences between groups were easily identified. Characteristics of each sequence group were summarized as follows.

- Group 1 as "Disengage-Deceleration", with a representative subsequence of "D1-DG" (AV deceleration followed by disengagement)
- Group 2 as "Hesitation", with a representative subsequence of "S1-A1-S1" (AV stops, proceeds, and stops again)
- Group 3 as "Stop", with a representative subsequence of "S1-X21" (AV hit by a second party after it stops)
- Group 4 as "Yield", with a representative subsequence of "Y-X21" (AV hit by a second party after it yields to the second party or a third party)
- Group 5 as "Hit Others", with representative subsequences of "DG-X1O" and "DG-X12" (AV disengagement followed by AV hitting a second party)
- Group 6 as "Left Turn", with a representative subsequence of "L1-X21" (AV was hit while making left turn)
- Group 7 as "Moving-Unexpected", with a representative subsequence of "A1-X21" (AV was hit while proceeding/accelerating)

Table 9  Clusters of AV crash sequences

| "Disengage-Deceleration" Group 1 | Count | "Hesitation" Group 2 | Count | "Stop" Group 3 | Count |
|---|---|---|---|---|---|
| D1-DG-X21 | 6 | S1-A1-S1-X21 | 4 | S1-X21 | 38 |
| D1-X21 | 4 | S1-A1-X21 | 4 | S1-PR2-X21 | 5 |
| ML2-DG-X21 | 3 | S1-A1-D1-X21 | 1 | S1-ML2-X21 | 3 |
| DT-D1-DG-V2-X21 | 2 | S1-A1-S1-A1-X21 | 1 | D1-S1-L2-X21 | 1 |
| ML3-D1-DG-X21 | 2 | S1-A1-S2-A2-X21 | 1 | D1-S1-X21 | 1 |
| MR3-D1-X21 | 2 | S1-A1-SA2-X32-X21 | 1 | PL1-PL2-S1-X21 | 1 |
| D1-DG-L2-X21 | 1 | S1-S2-A1-A2-X21 | 1 | R1-S1-XO1 | 1 |
| D1-ML2-X21 | 1 | *Total* | *13 (8%)* | S1-DG-X21 | 1 |
| D1-MR1-DG-X21 | 1 | | | S1-L1-X21 | 1 |
| D1-PL2-DG-X21 | 1 | | | S1-ML1-X12 | 1 |
| D1-PL2-X21 | 1 | | | S1-PL2-X21 | 1 |
| DG-B1-B2-SA2-X21 | 1 | | | S1-R1-X21 | 1 |
| L1-D1-DG-X21 | 1 | | | S1-R2-X21 | 1 |
| ML1-D1-DG-ML2-X21 | 1 | | | S1-XO1 | 1 |
| ML1-MR3-DG-MR1-X21 | 1 | | | *Total* | *57 (34%)* |
| ML3-D1-X21 | 1 | | | | |
| ML3-DG-X21 | 1 | | | | |
| MR2-DG-D1-X21 | 1 | | | | |
| MR3-DG-MR1-ML2-X21 | 1 | | | | |
| PL2-MR2-DG-X21 | 1 | | | | |
| PL3-D1-X21 | 1 | | | | |
| V2-D1-DG-X21 | 1 | | | | |
| *Total* | *35 (21%)* | | | | |
| **"Yield" Group 4** | **Count** | **"Hit Others" Group 5** | **Count** | **"Moving-Unexpected" Group 7** | **Count** |
| Y-X21 | 8 | DG-X1O | 2 | A1-X21 | 6 |
| R1-Y-X21 | 2 | A1-V2-DG-X12 | 1 | A1-MR2-X21 | 3 |
| Y-DG-X21 | 2 | ML1-MR1-DG-X12 | 1 | R1-X21 | 3 |
| Y-DG-D1-MR2-X21 | 1 | ML2-DG-X12 | 1 | A1-L2-X21 | 2 |
| Y-DG-ML2-X21 | 1 | ML3-DG-ML1-X12 | 1 | A1-ML2-X21 | 2 |
| Y-PL2-X21 | 1 | MR2-MR1-DG-X1O | 1 | A1-PL2-X21 | 2 |
| *Total* | *15 (9%)* | MR3-DG-MR1-X12 | 1 | R1-ML2-X21 | 2 |
| | | PL2-MR2-D2-DG-X12 | 1 | A1-ML2-DG-MR2-X21 | 1 |
| | | *Total* | *9 (5%)* | A1-PL2-DG-X21 | 1 |
| | | | | A1-PL2-MR2-X21 | 1 |
| | | | | A1-PR2-X21 | 1 |
| | | **"Left Turn" Group 6** | **Count** | A1-PR3-ML3-S1-X21 | 1 |
| | | L1-L2-X21 | 4 | A1-R2-X21 | 1 |
| | | L1-X21 | 2 | A1-V2-X21 | 1 |
| | | L1-L2-PL2-MR2-X21 | 1 | A1-XO1 | 1 |
| | | L1-L3-MR3-D1-X21 | 1 | L1-A1-X21 | 1 |
| | | *Total* | *8 (5%)* | ML1-D3-MR1-PR2-X21 | 1 |
| | | | | PL2-MR2-X21 | 1 |
| | | | | *Total* | *31 (18%)* |

Comparing the sizes, Group 1 consists of 35 sequences (21% of all 168 sequences), Group 2 has 13 (8%), Group 3 has 57 (34%), Group 4 has 15 (9%), Group 5 has 9 (5%), Group 6 has 8 (5%), and Group 7 has 31 (18%) sequences. Disengagements appeared concentratedly in Groups 1, 4, and 5. In Group 1's 35 crash sequences, disengagement appeared in 25 sequences. All disengagements following AV's yielding action were clustered in Group 4. All Group 5 sequences consisted of a disengagement event before AV colliding into an object or a second-party road user. In terms of other types of actions or maneuvers, stopping was mostly seen in Groups 2 and 3; yielding was seen in Group 4; merging and passing actions were mostly seen in Groups 1, 5, and 7; and left turning action was mostly seen in Group 6. Graph illustrations of the seven sequence patterns are shown in Figure 4. There were multiple different types of second-party road users involved in the 168 crash sequences, but to compactly present the sequence patterns, a motor vehicle was used in the illustrations to represent all types of second-party road users. For each sequence pattern, three panels of figures were used to illustrate the chronologically ordered events.

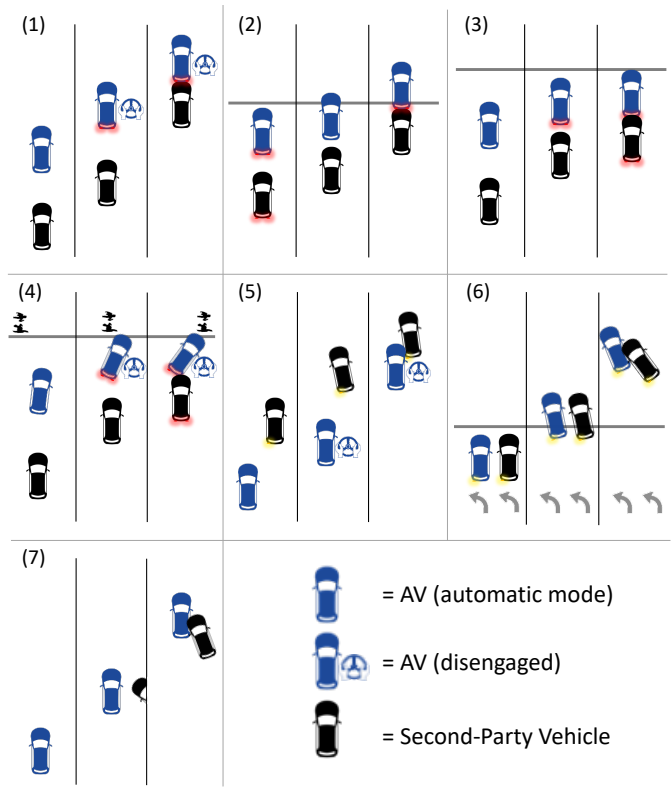

**Figure 4  Graph illustrations of AV crash sequence patterns**

**5.4    Cross-tabulation between Sequence Group and Other Variables**

A crash event sequence presents the progression of scenes with interactions between an AV and its surrounding moving objects (Ulbrich et al., 2015). In addition to moving object dynamics provided by crash sequences, multiple variables such as weather, lighting, road surface, road geometries, traffic control, and traffic conditions, need to be considered in developing test scenarios for AVs (Nitsche et al., 2017; Sander and Lubbe, 2018). Cross-tabulation analysis and Chi-square tests were carried out between sequence groups and several other variables that describe crash outcomes and environmental conditions. The purpose of cross-tabulation analysis is to evaluate the association between sequence groups and those other variables.

Figure 5 and Figure 6 illustrate results from a cross-tabulation between sequence group and two crash outcome measures, crash severity and manner of collision, respectively. The results showed that some Group 1, 2, 3, 6, and 7 sequences led to injuries. Comparing the distribution of crash severity in sequence groups, we found that Group 1 had the highest proportion (20%) of crash sequences that ended with injuries. Groups 6 and 7 both had 13% of crash sequences that ended with injuries. Group 3 had 12% of crash sequences that ended with injuries. The Chi-square test result (Chi-squared = 5.69, p-value = 0.47) showed that there is no significant association between sequence group and crash severity. However, after regrouping the sequence groups, with Groups 1, 2, 3, 6, and 7 in a new group, and Groups 4 and 5 in another new group, the Chi-square test result (Chi-squared = 3.78, p-value = 0.08) showed that there is a more significant association between new sequence groups and crash severity.

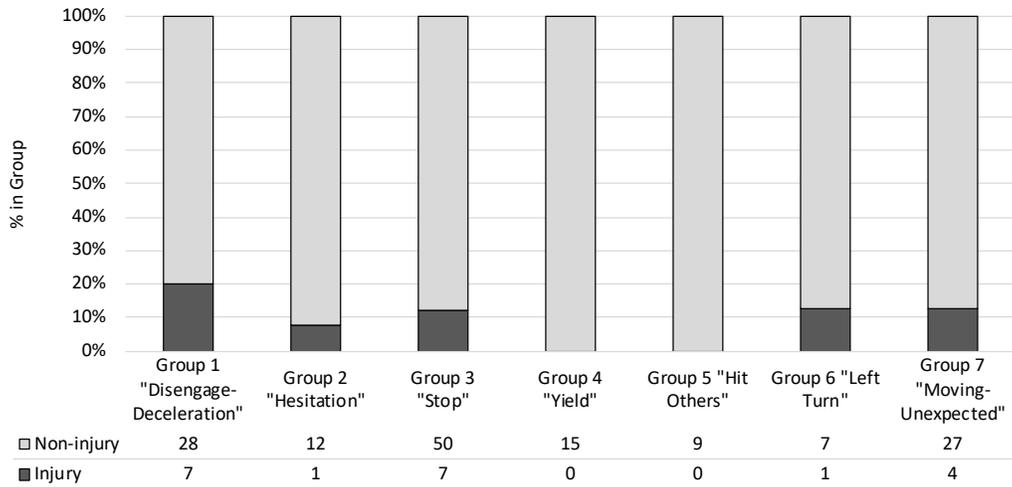

**Figure 5  Crash severity distribution by sequence group**

Manners of collision distributed differently across different sequence groups. Rear end, sideswipe, and broadside are the three most frequent manners of collision. Sequences in Groups 1, 3, 5, and 7 led to a larger variety of manners of collision, with 4-5 types in each group. A Chi-square test between manners of collision and sequence groups showed a significant association (Chi-squared = 83.73, p-value = 0.00). Different sequence groups led to different compositions of collision manners.

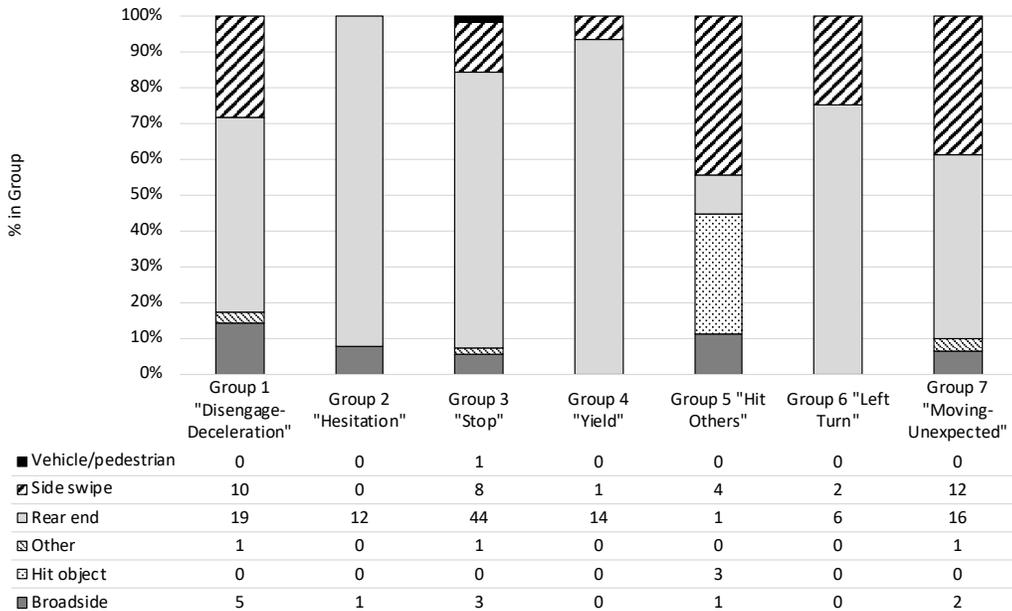

**Figure 6  Manner of collision distribution by sequence group**

Figure 7 and Figure 8 illustrate results from cross-tabulations between crash sequence groups and variables describing environmental conditions including facility type and time of day. Crash sequence groups distributed differently across facility types. The intersections (including ramp terminals) category had the largest variety of sequence groups. Groups 2 and 6 sequences only took place at intersections. The Chi-square test result (Chi-squared = 43.86, p-value = 0.00) confirmed that there is a significant difference in sequence group distribution across facility type.

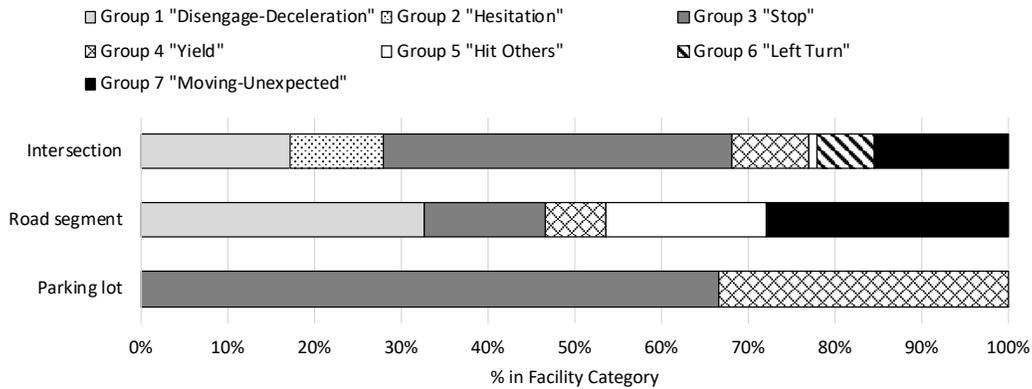

**Figure 7  Sequence group distribution by facility type**

Time of day is a variable closely related to weather, lighting, and traffic conditions. Based on a visual check, the three most frequently observed AV crash sequence groups were Groups 1, 3, and 7 for both daytime and nighttime. Group 2 crash sequences were only observed during daytime but not nighttime. A Chi-square test result (Chi-squared = 9.22, p-value = 0.16) did not show a significant difference in sequence group distributions between daytime and nighttime.

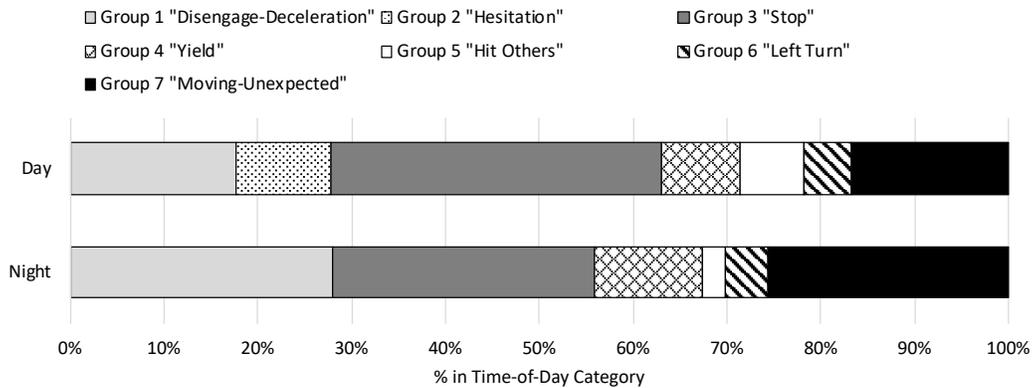

**Figure 8  Sequence group distribution by time of day**

As AV technology is rapidly developing, AV crash sequence patterns are expected to change over the years. Therefore, for the consideration of designing AV test scenarios, a cross-tabulation was carried out between sequence groups and the years, with results illustrated in Figure 9. Based on a visual check, the distributions of sequence groups varied across the years from 2015 to 2019. A Chi-square test result (Chi-squared = 42.40, p-value = 0.01) showed a significant difference of sequence group distribution across the years.

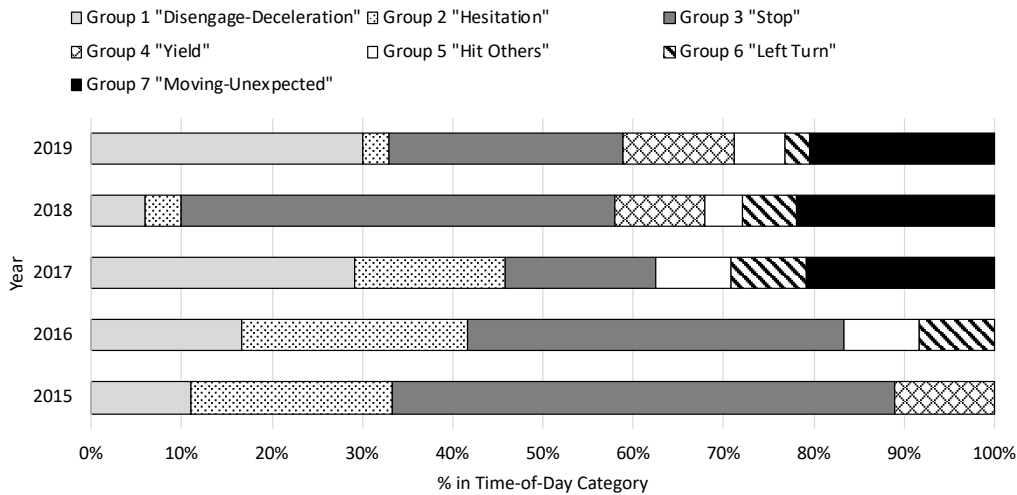

Figure 9  Sequence group distribution by year

## 6   Discussion

With an analysis of California AV crash sequences, we summarized the most frequent events and subsequences, estimated transition probabilities between events, identified cohort groups of sequences, and evaluated the association between sequence groups and variables measuring crash outcome and environmental conditions. The analysis results led to the following findings:

- The most representative subsequence of California AV crashes was "collision following AV stop".
- Disengagements were observed in 24% of AV crash sequences. Disengagements were mostly initiated due to operator precaution and detection of other road users' reckless behavior. Disengagements in the studied AV crash sequences were mostly followed by an immediate collision with other road users or objects, not leaving enough time for the human operator to take further actions.
- AV crash sequences were clustered into seven groups. Each sequence group has a representative subsequence, presenting unique characteristics of AV crash progression.
- AV crash sequence groups were significantly associated with variables measuring crash outcomes and describing environmental conditions, indicating that scenarios described by combinations of event sequences and environmental condition variables can lead to various crash outcomes.

The sequence analysis results show that there are patterns in AV crash sequences, which provide information about AV crash progression and form distinctive cohort groups. Sequence groups were shown to lead to different crash outcomes and were associated with environmental condition variables. Events such as disengagement, with its preceding and succeeding events, are unique to AV operations and worthy of consideration in designing AV test scenarios. Also, AV crash sequence patterns changed across the years. As AV technology develops, new crash sequence patterns should be accommodated in AV testing.

A scenario-based AV safety testing framework was developed with sequence of events embedded as a core component. Figure 10 provides a simplified illustration of the framework. A more comprehensive framework can be built based on this one with additional details. This framework consists of a test scenario setup module and a performance evaluation module, which are both built around sequence of events. According to Koopman and Fratrik, AV safety evaluation should be able to validate factors from a four-dimensional validation space with the axes of {Operational design domain (ODD), Object and event detection and response (OEDR), Maneuvers, Fault Management} (Koopman and Fratrik, 2019). Our proposed framework captures all these factors through modeling actions and interactions with sequence of events.

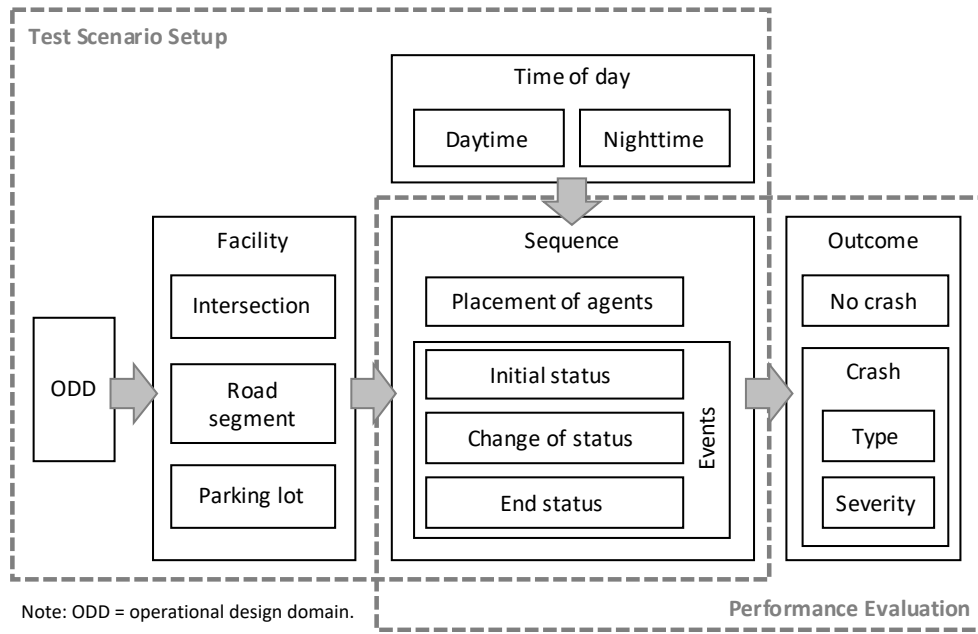

**Figure 10  AV safety testing framework with sequence of events embedded**

An ODD describes the specific domain in which an automated driving system is designed to properly operate. With a determined ODD, variables such as types of roadways, ranges of speed, time of day, and limits of weather are determined (NHTSA, 2017). Within the limits of a certain ODD, AV test scenarios can be developed. A scenario is described by variables including road geometries, roadside objects and rules, temporary modifications and events, moving objects, natural environmental conditions, and digital information (Sauerbier et al., 2019). Crash sequences can be used to encode interactions between moving objects. The crash sequence patterns generated from this study provide a semantic-level description of moving object interactions. Together with the roadway features and environmental condition variables, crash sequence patterns form the basic structure of a test scenario. On the foundation of semantic-level scenarios, concrete scenarios can be generated by defining parameter ranges and specific values for each event and action in the sequences, as well as for each roadway and environmental condition variables (Menzel et al., 2019, 2018). Such a process requires additional microscopic AV operations and incidents data, which can be collected from AV field operational tests or NDS databases. To ensure efficient data collection, crash sequence patterns such as the set found in this study, can also be used as a guideline to identify frequent and rare crash cases for data collection.

The following is a procedure to set up test scenarios and evaluate AV safety performance using the proposed framework:

- Based on a determined ODD, a type of facility of interest is selected as the base environment of a test scenario. As many design characteristics of the facility should be considered as possible.
- Environmental condition factors such as time of day, lighting, and weather, are set up on the basis of the selected facility, to create various conditions for testing.
- Moving objects (AVs and other road users) are deployed in the test environment. Representative sequence patterns obtained from sequence analysis of historical crashes can be used to guide the setup of where and when the moving objects appear in the test environment, as well as the interactions between moving objects.
- The moving objects then interact in the test environment and generate crash outcomes, measured by variables such as crash rate, manner of collision, or injury severity. Surrogate safety measures for conflicts are alternative options to describe outcomes. After repeated tests, AVs' safety performance is evaluated based on the test-generated crash/safety outcomes.

# 7 Conclusions

As AV development and testing expand, safety evaluation of such vehicles needs to catch up. Through the analysis of 168 AV crash sequences, this research identified patterns in AV crash sequences, which led to a discussion on potential uses of crash sequences in AV safety testing. The conclusion is that crash sequence patterns capture the characteristics of AV crash progression and should be useful in generating AV test scenarios.

Compared with previous studies on exploring California AV crash and disengagement patterns, this study investigated AV crashes and disengagements from a different perspective, sequence of events leading to a crash. Patterns in crash sequences were identified, with AV crash sequences clustering into seven distinctive cohort groups. Cross-tabulation analysis showed that sequence groups are significantly associated with variables measuring crash outcomes and describing environmental conditions. AV crash sequences can be used in generating semantic-level AV test scenarios. Based on the findings, an AV safety testing framework was proposed with sequence of events embedded as a core component.

In addition to the contribution in discovering AV crash sequence patterns, this study showed the value of crash sequence analysis, and reemphasized the importance of collecting crash sequence data. Although the importance of crash sequence of events was stressed by NTSB, reporting such information was not required by the California DMV, nor were any guidelines provided for including crash sequence information in text narratives (California DMV, 2018; NTSB, 2011). Crash sequence information was buried in the narratives of AV crash reports. In addition to descriptive summary of crash report data, this study carried out a more in-depth analysis, which helped us discover more informative patterns in AV crashes than two very recent studies using the same data source of crash report text narratives (Alambeigi et al., 2020; Das et al., 2020). With crash sequences, we were also able to further analyze the relationship between AV disengagements and crashes, and better understand the role of disengagement in crashes that happened during AV field tests. Recent studies of AV disengagements focused on finding contributing factors to disengagements rather than evaluating the connection between disengagements and crashes (Boggs et al., 2020a; Wang and Li, 2019a).

Limitations of this study are primarily in the use of crash reports filled out by different AV-testing organizations and submitted to the California DMV. One author classified the events and developed the event sequences to maintain consistency. The recording of events in crash sequences was based on crash text narratives and researchers' comprehension of such narratives. Consistency in crash sequences was enhanced through a two-phase encoding process and by having one author perform this task.

For future work, a similar analysis will be carried out using more AV crash data as they become available. Improved encoding and sequence analysis methods will also be used. With crash sequence data available in historical human-driven vehicle crash databases, a comparative study will be carried out between patterns in AV crash sequences and human-driven vehicle crash sequences. The authors strongly recommend that federal and state transportation agencies require AV testing organizations to share microscopic, event-level data of AV disengagements and crashes that occur during public-road tests and make the data available to safety researchers. Detailed data would enable a much more informed AV testing and evaluation process, bring transparency to public-road AV testing, and enhance public trust in AVs.

**Acknowledgements**

The authors would like to thank Adam Francour and Beau Burdett of the University of Wisconsin Traffic Operations and Safety Laboratory for their help in proofreading the manuscript and providing valuable suggestions.

**Funding**

This study was sponsored by the Safety Research using Simulation University Transportation Center (SAFER-SIM). SAFER-SIM is funded by a grant from the U.S. Department of Transportation's University Transportation Centers Program (69A3551747131). The work presented in this paper remains the responsibility of the authors.